# Dendrite Suppression by Shock Electrodeposition in Charged Porous Media

*Ji-Hyung Han[1], Miao Wang[1], Peng Bai[1], Fikile R. Brushett[1], and Martin Z. Bazant[1,2]\**

Department of Chemical Engineering[1] and Department of Mathematics[2],
Massachusetts Institute of Technology, Cambridge, MA 02139, USA

**ABSTRACT**

It is shown that surface conduction can stabilize electrodeposition in random, charged porous media at high rates, above the diffusion-limited current. After linear sweep voltammetry and impedance spectroscopy, copper electrodeposits are visualized by scanning electron microscopy and energy dispersive spectroscopy in two different porous separators (cellulose nitrate, polyethylene), whose surfaces are modified by layer-by-layer deposition of positive or negative charged polyelectrolytes. Above the limiting current, surface conduction inhibits growth in the positive separators and produces irregular dendrites, while it enhances growth and suppresses dendrites behind a deionization shock in the negative separators, also leading to improved cycle life. The discovery of stable uniform growth in the random media differs from the non-uniform growth observed in parallel nanopores and cannot be explained by classic quasi-steady "leaky membrane" models, which always predict instability and dendritic growth. Instead, the experimental results suggest that transient electro-diffusion in random porous media imparts the stability of a deionization shock to the growing metal interface behind it. Shock electrodeposition could be exploited to enhance the cycle life and recharging rate of metal batteries or to accelerate the fabrication of metal matrix composite coatings.

*Corresponding author: M.Z. Bazant (bazant@mit.edu)

**INTRODUCTION**

Pattern formation by electrodeposition has fascinated scientists in recent decades since Brady and Ball[1] first attributed the mechanism of copper dendritic growth to diffusion-limited aggregation[2]. It was later discovered that morphology selection is also influenced by electromigration and convection in free solutions[3-10]. Here, we report some surprising effects of electromigration on electrodeposition in weakly charged porous media, including the possibility of stabilizing the growth and eliminating dendrites at high rates.

Suppressing dendrites in porous separators is a critical challenge for high-energy-density batteries with Li[11,12], Zn[13], Na, Cd or other metal anodes. Dendrites accelerate capacity fade and cause dangerous short circuits[11,12]. Dendrites can be blocked by stiff, dense separators[14,15], but usually only at the cost of large internal resistance. Another strategy is to manipulate ionic fluxes near the anode via competing side reactions that interfere with electrodeposition at protrusions[16-19] or enhance surface diffusion[64]. More stable metal cycling has also been demonstrated by altering the separator chemistry, e.g. with lithium-halide salts[64], nanoparticles with tethered ionic-liquid anions[20], hydrophilic separators[21], electrolytes with large anions[22], and certain solid polymer electrolytes[64,65].

Motivated by reducing space charge[23,24], several studies have shown that supplying extra anions by charged nanoparticle dispersion[25] or solvent-in-salt electrolyte[26] can improve battery cycling, although dendrites were not visualized. However, according to theory[32] and experiments on dendritic growth[8] and electrodialysis[30,33,34], it is unlikely that extended space charge ever forms in free solutions. In the case of copper electrodeposition, morphological instability occurs immediately upon salt depletion at the cathode, which enhances ionic flux to the tips, avoid space charge, and preserves thin double layers[7,16,35]. At the same time, the Rubinstein-Zaltzman hydrodynamic instability can lead to vortices that sustain over-limiting current (OLC), faster than electro-diffusion[30,33,34,36,37]. This phenomenon is well established in electrodialysis[30,33,34,37-40] and nanofluidics[41] and may also explain electroconvection observed around dendrite tips[3,8].

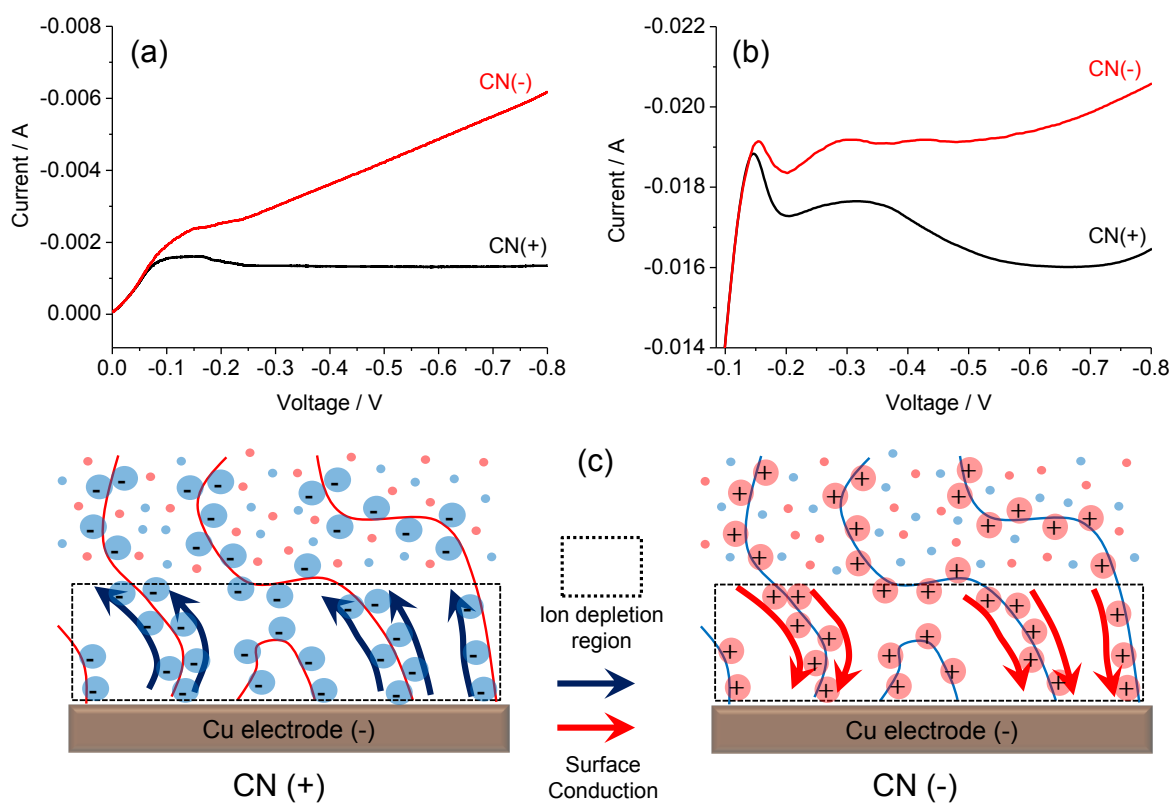

**Figure 1.** Voltammetry of positively and negatively modified cellulose nitrate (CN) membranes of exposed area 1.327 cm$^2$ between Cu electrodes in (a) 10 mM CuSO$_4$ at 1 mV/s and (b) 100 mM CuSO$_4$ at 10 mV/s. (c) Physical pictures of the effect of surface conduction on electrodeposition in a charged random porous media, driven by large electric fields in the ion depletion region.

Here, we establish new principles of morphology control for electrodeposition in porous media. By exploiting the physics of deionization shock waves[43], we show that porous separators with thin electric double layers ("leaky membranes"[27]) can either stabilize or destabilize metal electrodeposition at high rates, depending on the sign of their surface charge. Our initial model system is a symmetric copper cell consisting of a porous cellulose nitrate (CN) or polyethylene (PE) separator with positive or negative polyelectrolyte coatings, which is compressed between two flat copper electrodes in copper sulfate solutions. The current-voltage relations in both cases (Fig. 1a and 1b) show common plateaus around the diffusion-limited current because surface conduction is negligible compared to bulk electro-diffusion. At higher voltages, however, strong salt depletion occurs at the cathode, and dramatic effects of the surface charge are observed (Fig. 1c). The positive separator exhibits reduced cation flux, opposed by surface conduction[28], while the negative separator exhibits OLC sustained by surface conduction[7,27,28], which also leads to a transient deionization shock[29-31 43] ahead of the growth.

We have discovered that the interaction between these nonlinear transport phenomena and the growing deposit is strongly dependent on the porous microstructure, as shown in Fig. 2. In a recent publication[28], we showed that surface conduction can profoundly influence the *pore-scale* morphology of copper growth in ordered anodic alumina oxide (AAO) membranes. In such materials with *non-intersecting* parallel nanopores, diffusion-limited metal growth is inherently non-uniform and leads to a "race of nanowires"[62]. Above the limiting current, there is a transition to new non-uniform growth modes, either nanotubes following separate deionization shock waves in each pore of the negatively-charged membrane (Fig. 2b) or slowly penetrating, pore-center dendrites in the positively-charged membrane (Fig. 2a). Here, we demonstrate nearly opposite effects of surface conduction on the *electrode-scale* morphology in random CN membranes with *well-connected* pore networks. Above the limiting current, some low-density dendritic structures penetrate into the positive membrane (Fig. 2c), but, remarkably, the growth is uniform, dense, and reversible in the negative membrane, which we attribute to the propagation of a single flat, stable deionization shock ahead of the deposit (Fig. 2d).

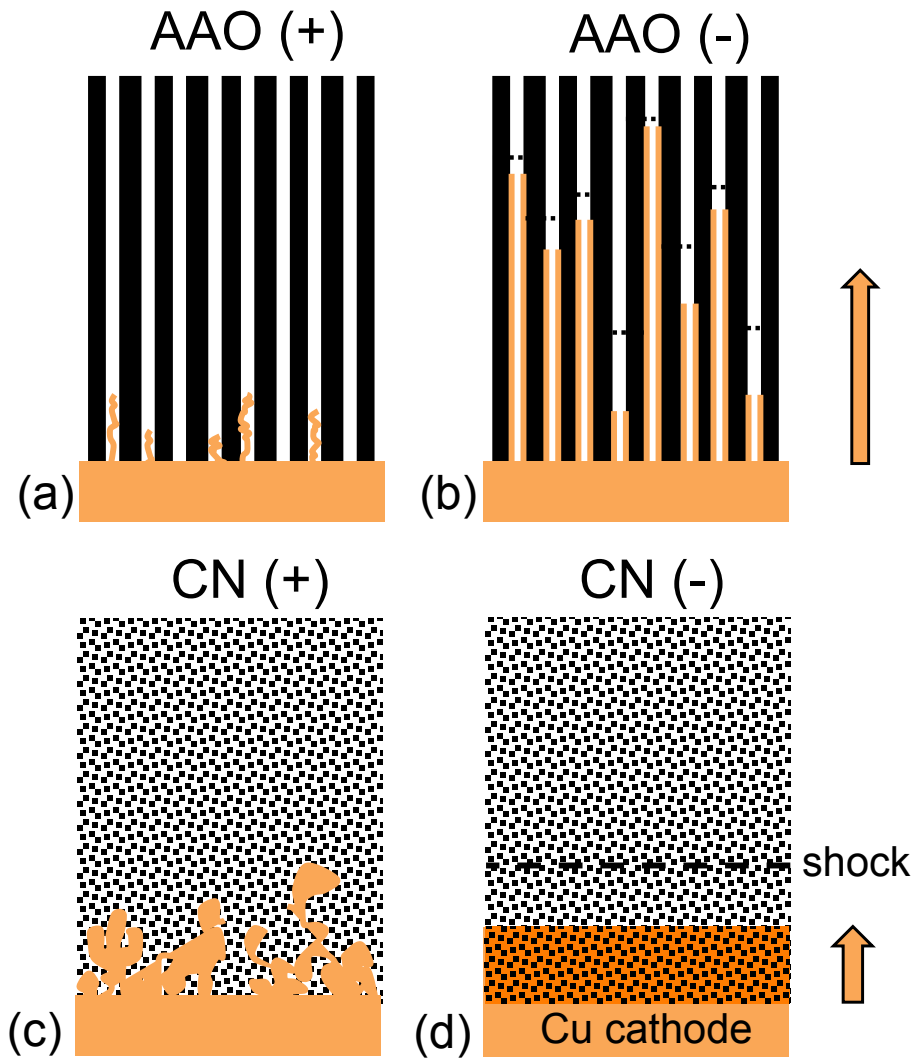

**Figure 2.** Morphology selection principles for fast electrodeposition (exceeding diffusion limitation) in charged, porous media with ordered pores (e.g. anodic aluminum oxide, AAO, from our previous work[28]) versus random pores (e.g. cellulose nitrate, CN, from this work). In parallel nanopores, (a) positive surface charge suppresses metal penetration or allows thin dendrites avoiding the pore walls, while (b) negative charge promotes non-uniform surface coverage leading to metal nanotubes of different lengths growing behind deionization shock waves (dashed lines). In well-connected, random nanopores, (c) positive surface charge blocks penetration or allows low-density porous dendrites, while (d) negative charge leads to a flat metal-matrix composite film, stabilized by a macroscopic shock wave propagating ahead of the growth.

**THEORY**

In porous media, the physical mechanisms for OLC are very different from those in free solutions and just beginning to be explored. According to theory[7], supported by recent microfluidic experiments[31], if the counterions (opposite to the pore surface charge) are the ones being removed, then extended space charge is suppressed, and electro-osmotic instability is replaced by two new mechanisms for OLC: (1) surface conduction by electromigration, which dominates in submicron pores[27,28], and (2) surface convection by electro-osmotic flow, which dominates in micron-scale pores[27,32,40,42]. Regardless of whether OLC is sustained by constant current[29,43] or constant voltage[44], the ion concentration profile develops an approximate discontinuity that propagates into the porous medium, leaving highly deionized fluid in its wake, until it relaxes to a steady linear profile in a finite porous slab[27,45]. This "deionization shock wave"[43] is analogous to concentration shocks in chromatography, pressure shock waves in gases, stop-and-go traffic, glaciers, and other nonlinear kinematic waves[46].

The influence of surface conduction on electrodeposition was recently discovered in our investigations of copper electrodeposition in AAO membranes with modified surface charge[28]. Below the limiting current, surface conduction is negligible if the double layers are thin (small Dukhin number), but surface conduction profoundly affects the growth at high currents. With positive surface charge, growth is blocked at the limiting current by oppositely-directed surface conduction (electro-migration) and surface convection (electro-osmotic flow); above a critical voltage, some dendrites are observed avoiding the pore walls, likely fed by vortices of reverse electro-osmotic flow returning along the pore centers. With negative surface charge, the growth is enhanced by surface conduction until the same critical voltage, when surface dendrites and ultimately smooth surface films grow rapidly along the walls. These phenomena are consistent with the theory of OLC in a single microchannel[7,31], but we expect different behavior in random media with interconnected pores.

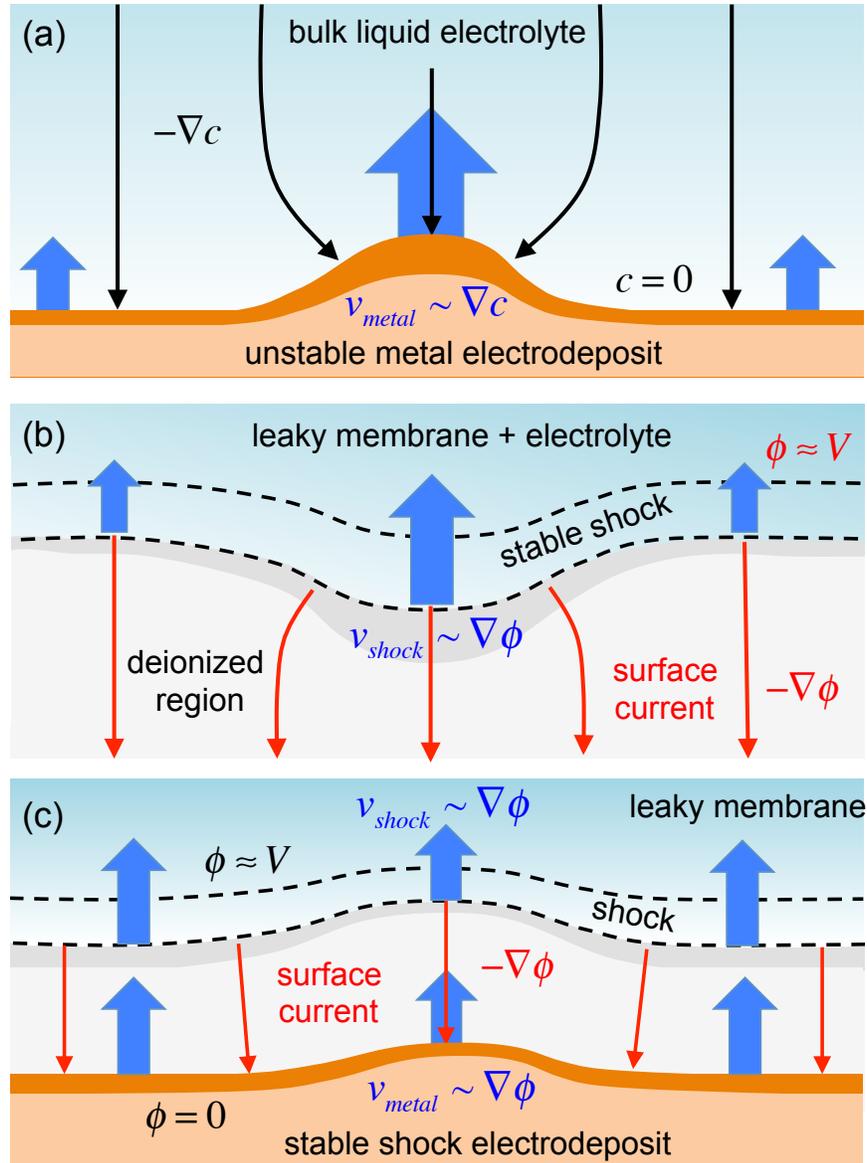

**Figure 3.** Basic physics of shock electrodeposition. (a) Dendritic instability of electrodeposition in free solution. (b) Stability of deionization shock propagation in a leaky membrane. (c) Stabilization of electrodeposition behind a deionization shock.

The motivation for our experiments is the theoretical prediction that a flat deionization shockwave is nonlinearly stable to shape perturbations[43], since we hypothesize that this stability could be imparted to an electrodeposit growing behind a propagating shock. In free solution, dendritic growth occurs soon after salt depletion, owing to the simple fact that a surface protrusion receives more flux, thereby causing it to protrude further[1] (Fig. 3a). This is the fundamental instability mechanism of Laplacian growth, which leads to fractal patterns by continuous and deterministic viscous fingering[47] or discrete and stochastic diffusion-limited aggregation (DLA)[2]. In contrast, the propagation of deionization shockwaves is controlled "from behind" by the high resistance of the ion depletion zone. As shown in Fig. 3b, a lagging region of the shock will have more flux leaving by surface conduction, causing it to advance back to the stable flat shape. The dynamics of a thin shock is thus equivalent to Laplacian dissolution[43], the stable time reversal of Laplacian growth[48], and this suggests that transport-limited electrochemical processes occurring behind the shock might proceed more uniformly as well.

What would happen if a stable deionization shock precedes an unstable growing electrodeposit in a charged porous medium? According to the simplest theoretical description, the classical[49] "leaky membrane" model[27,43,45] (LMM), the answer depends on the importance of transient diffusion ahead of the shock. The ion concentrations $c_i(\vec{x},t)$ and electrostatic potential $\phi(\vec{x},t)$ satisfy the Nernst-Planck equations,

$$\frac{\partial c_i}{\partial t} + \vec{u}\cdot\nabla c_i = -\nabla\cdot\vec{F}_i = \nabla\cdot\left(D_i\nabla c_i + z_i e M_i c_i \nabla\phi\right) \tag{1}$$

and macroscopic electroneutrality,

$$\sum_i z_i e c_i + \rho_s = 0 \tag{2}$$

including the surface charge density per volume, $\rho_s$. The mean flow is incompressible, driven by gradients in dynamical pressure, electrostatic potential, and chemical potential, respectively,

$$\nabla\cdot\vec{u} = 0, \qquad \vec{u} = -k_D\nabla p - k_{EO}\nabla\phi - k_{DO}\nabla\ln c_i. \tag{3}$$

The macroscopic ionic diffusivities, $D_i$, and mobilities, $M_i$, Darcy permeability, $k_D$, electro-osmotic mobility, $k_{EO}$, and diffusio-osmotic mobility, $k_{DO}$, depend on $c_i$ and $\phi$, but not on their gradients or (explicitly) on position. This approximation is reasonable for surface conduction in nanopores, but neglects hydrodynamic dispersion due to electro-osmotic flow in micron-sized pores[50] or pore network loops[40], for which no simple model is available[32,42]. Assuming a transport limited growth process, the moving electrode surface has Dirchlet ($c_i = \phi = 0$) and Neumann ($\hat{n}\cdot\vec{u} = 0$) boundary conditions.

With these general assumptions, we observe that the steady-state LMM, Eqs. (1)-(3), falls into Bazant's class of conformally invariant nonlinear partial differential equations[51]. The profound implication is that *quasi-steady* transport-limited growth in a leaky membrane (with growth velocity opposite to the active-ion flux, $\vec{v} \sim -\vec{F}_1$) is in the same universality class[52] as Laplacian growth[53,54] and thus always unstable. This explains the recent theoretical prediction that negative charge in a leaky membrane cannot stabilize quasi-steady electrodeposition, although it can reduce the growth rate of the instability[55], consistent with the improved cycle life of lithium batteries with tethered anions in the separator[20,25].

In contrast, copper electrodeposition experiments in free solution have shown that the salt concentration profile is *unsteady* prior to interfacial instability[35] and forms a "diffusive wave" ahead of growing dendrites[4-7] with the same asymptotic profile as a deionization shock[43]. In a negatively charged medium, before the salt concentration vanishes at Sand's time, the diffusion layer sharpens and propagates away from the electrode as deionization shock[27,45], which could perhaps lead to stable, uniform "shock electrodeposition" in its wake, as outlined in Fig. 3c. Since the LMM neglects many important processes, however, such as surface diffusion[56], surface convection[32,50,56], pore-scale heterogeneity[57], and electro-hydrodynamic dispersion[40,42,50], we turn to experiments to answer this question.

**EXPERIMENTAL RESULTS**

In order to isolate the effects of charged porous media, we use the same copper system ($Cu|CuSO_4|Cu$) studied over the past three decades by physicists, as a canonical example of diffusion-limited pattern formation[1,3]. Compared to lithium electrodeposition and electrodissolution, which involves complex side reactions related to the formation and evolution of the solid-electrolyte interphase (SEI), this system is simple enough to allow quantitative interpretation of voltammetry in nanopores[28] and microchannels[3,7,35]. A unique feature of our experiments is that we control the surface conductivity by modifying the separator surface charge by layer-by-layer (LBL) deposition of charged polymers. We also demonstrate the role of *pore connectivity* for the first time by choosing random porous media, such as cellulose nitrate (CN), with similar pore size (200~300 nm) as the parallel nanopores of AAO from our recent study that introduced this method[28]. We denote the charge-modified positive and negative membranes as CN(+) and CN(-), where excess sulfate ions ($SO_4^{2-}$) and cupric ions ($Cu^{2+}$), respectively, are the dominant counter-ions involved in surface conduction (Fig. 1c).

As noted above, voltammetry clearly shows the nonlinear effect of surface conduction. Fig. 1a shows current-voltage curves of CN(+) and CN(-) in 10 mM $CuSO_4$ at a scan rate of 1 mV/s, close to steady state. In the low-voltage regime of slow reactions[28] (below -0.07 V), the two curves overlap since the double layers are thin, and surface conduction can be neglected compared to bulk diffusion (small Dukhin number)[27,40]. At the diffusion-limited current, huge differences in CN(+) and CN(-) are suddenly observed. While the current in the CN(+) reaches -1.5 mA around -0.1 V and maintains a limiting current of -1.3 mA, the CN(-) shows a strong linear increase in current, i.e. constant over-limiting conductance. The data are consistent with the surface conduction (SC) mechanism (Fig. 1c), which is sensitive to the sign of surface charge[28,50]. With negative charge, $Cu^{2+}$ counter-ions provide surface conduction to "short circuit" the depletion region to maintain electrodeposition. With positive charge, the $SO_4^{2-}$ counter-ions migrate away from the cathode, further blocking $Cu^{2+}$ ions outside the depletion region in order to maintain neutrality. At higher salt concentration, 100 mM $CuSO_4$, sweeping at 10 mV/s, the results are similar (Fig. 1b) with no effect of SC below -0.15 V, limiting current of -19 mA for CN(+), and overlimiting conductance for CN(-), although the effect of SC is weaker (smaller Dukhin number), and transient current overshoot and oscillations are observed[28,58].

Striking effects of surface charge are also revealed by chronopotentiometry (Fig. 4). When constant OLC (-5 mA) is applied in 10 mM $CuSO_4$ solution, CN(+) exhibits large, random voltage fluctuations, which we attribute to the blocking of cation transport by the reverse SC of $SO_4^{2-}$ counter-ions near the cathode. Large electric fields drive unstable electro-osmotic flows, some dendritic growth, and water electrolysis, consistent with observed gas bubbles. Metal growth is mostly prevented from entering the CN(+) membrane, so it is easily separated from the cathode after the experiment. In stark contrast, CN(-) maintains low voltage around -100 mV, as expected since the SC of $Cu^{2+}$ counter-ions sustains electrodeposition under OLC regime. More importantly, the electrodeposited Cu film in CN(-) is perfectly uniform, as shown in the SEM image of Fig. 4(b), consistent with the theoretical motivation above, based on the stability of deionization shock propagation ahead of the growth.

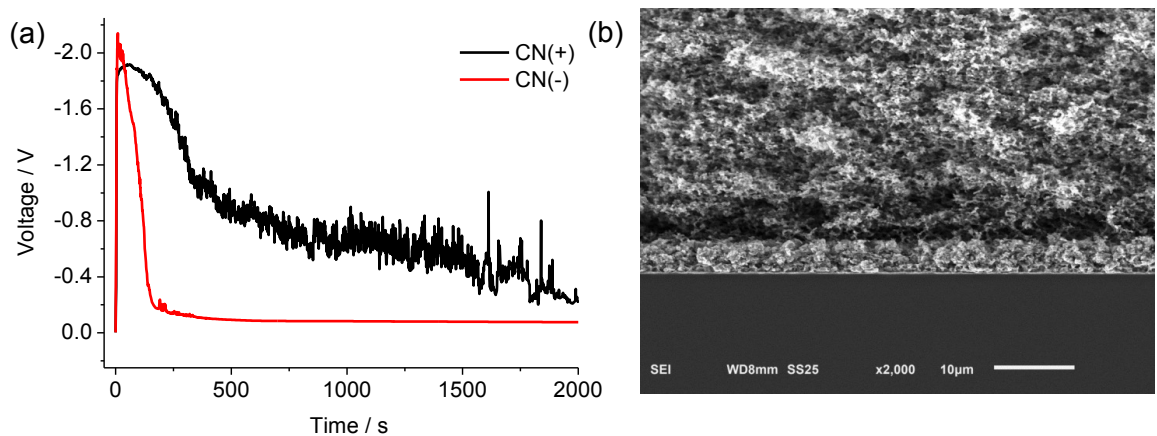

**Figure 4.** (a) Chronopotentiometry data for CN(+) and CN(−) membranes at −5 mA for 2000 s in 10 mM $CuSO_4$. (b) SEM image of a uniform Cu film in CN(−) grown by shock electrodeposition during OLC.

Figure 5 clearly shows the suppression of dendritic instability. When OLC (−20 mA) is applied in 100 mM $CuSO_4$ for 2000 s, irregular electrodeposits are generated in CN(+) (Fig. 5a). This imposed current exceeds the limiting current (−17 mA) measured by voltammetry (Fig. 1b), so the observed low-density stochastic growth, which is opposed by surface conduction, may result from vortices of surface electroconvection, driven in the reverse direction by huge electric field in the depletion region. Once again, under the same experimental conditions, we obtain a highly uniform Cu film in CN(−) (Fig. 5b) by shock electrodeposition.

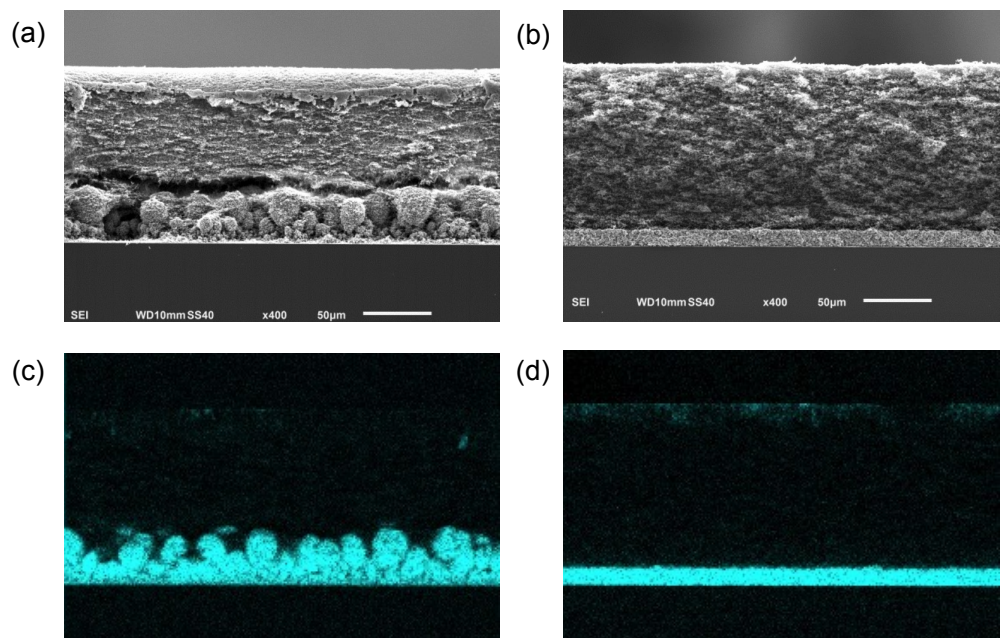

**Figure. 5.** Morphologies of Cu film depending on surface charge of CN membrane: CN(+) (a and c) and CN(-) (b and d). Cu electrodeposition is carried out in 100 mM $CuSO_4$ by applying -20 mA for 2000 s. SEM images (a and b) and EDS mapping analysis of Cu (c and d).

The difference in morphology of Cu electrodeposits between CN(+) and CN(-) can also be precisely confirmed by EDS mapping analysis of Cu element (Fig. 5c and 5d). The Cu film in CN(-) shows more compact and flat morphology, consistent with simple estimates of the metal density. Based on the applied current (-20 mA), nominal electrode area (1.0 cm × 1.5 cm) and time (2000 s), pure copper would reach a thickness of 19.6 µm, which would be increased by porosity, but also lowered by fringe currents, side reactions, and metal growth underneath the membrane. The penetration of copper dendrites in CN(+) to a mean distance of 45 µm, supports the direct observation of low density ramified deposits, while the smaller penetration, 12.8 µm, into CN(-) suggests that shock electrodeposits densely fill the pores.

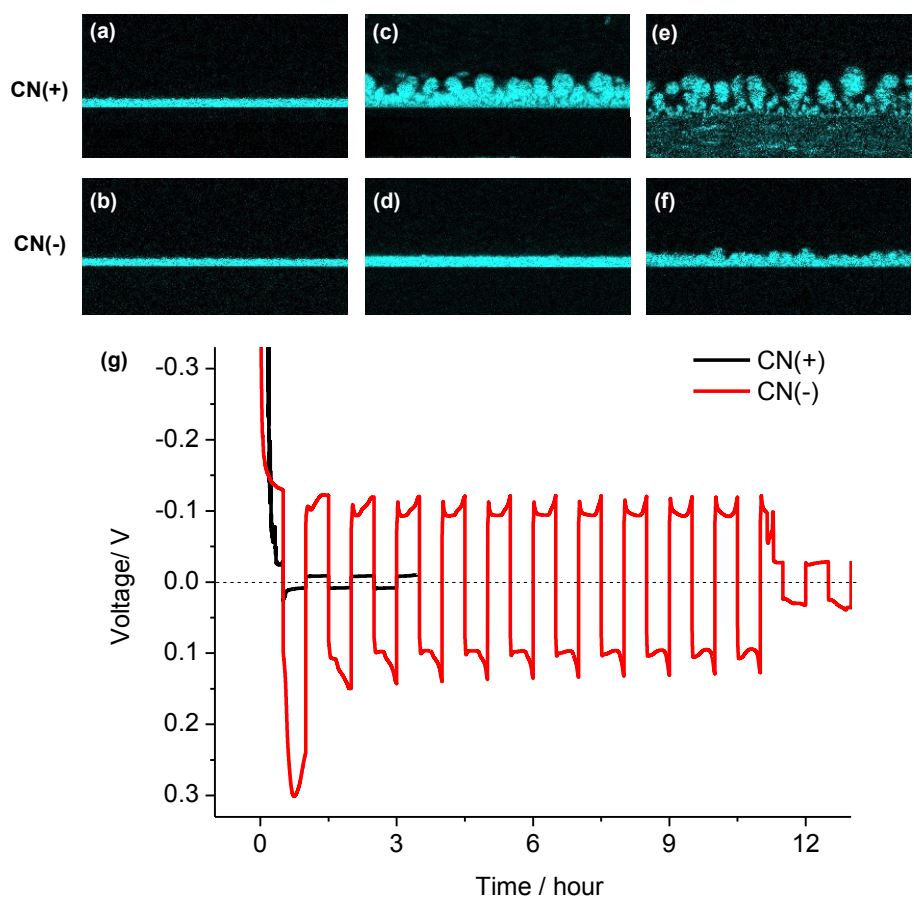

**Figure 6**. EDS mapping analysis of Cu element. Cu is electrodeposited in CN(+) (a, c, e) and CN(-) (b, d, f) membranes at constant current densities in 100 mM $CuSO_4$ for 2000s (a, b) -15 mA, (c, d) -20 mA, and (e, f) -25 mA. (g) Galvanostatic cycling profiles of CN(+) and CN(-) using a symmetric copper cell: Cu is electrodeposited and electrodissolved under extreme OLC (25 mA) for 1800 s in 100 mM $CuSO_4$.

The variation of morphology with applied current is demonstrated in Fig. 6. For under-limiting current (-15 mA), both cases exhibit a uniform Cu film (Fig. 6a and 6b), independent of surface charge, as expected when surface conduction is weak compared to bulk electro-diffusion within the pores (small Dukhin number). As the applied current is increased, highly irregular, dendritic electrodeposits are generated in CN(+). When extreme OLC (-25 mA) is applied, CN(+) shows much less dense dendritic growth, and weak adhesion of the membrane to the cathode leading to its peeling off (Fig. 6e). On the other hand, shock electrodeposition in CN(-) suppresses dendritic growth and produces uniform, dense Cu films, which show signs of instability only at very high currents (Fig. 6f).

The observed morphologies shed light on the different cycling behavior for positive and negative membranes under extreme currents (± 25 mA), as shown in Fig. 6g. The unstable dendritic growth of CN(+) results in short-circuit paths that cause the voltage to drop quickly to 5 mV in the first cycle. Although further cycles are possible, the voltage never recovers. In contrast, the more uniform growth observed in CN(-) is associated with stable cycling around ± 100 mV, in spite of the large nominal current density (± 18.8 mA/cm$^2$), well above the limiting current. After eleven cycles the voltage drops to 30 mV, but further cycling is still possible without short circuits. Improved cycling life has also recently been reported for lithium metal anodes with separators having tethered anions[59], albeit at much lower currents (0.5 mA/cm$^2$) without observing the deposits. Our observation of stable shock electrodeposition may thus have broad applicability, including rechargeable metal batteries.

In order to investigate the generality of this phenomenon and its potential application to batteries, we repeated the same experimental procedures for several commercially available, porous polymeric battery separators. Here, we report results for a 20 $\mu$m thick Celgard K2045 polyethylene (PE) membrane with a pore size of 50 nm, porosity of 47%, and a tortuosity of 1.5, which was modified using the same layer-by-layer (LBL) assembly sequence for either positively or negatively charged membrane. As is evident in the voltammetry of (+) and (-) PE membranes (Fig. 7a), similar OLC behavior, consistent with the nonlinear effect of surface conduction, is observed as the copper electrode is polarized at a scan rate of 2 mV/s in 100 mM CuSO$_4$ solution. Additional data for 10 mM solutions can be found in the Supporting Information. Once diffusion limitation begins to dominate at approximately -0.15 V, consistent discrepancies in the current-voltage curve can again be attributed to surface conduction, which enhances Cu$^{2+}$ transport in the PE(-) membrane, as anions (SO$_4^{2-}$) in the double layer of the PE(+) membrane further block the transport of Cu$^{2+}$ inside the depleted region near the cathode. The sudden increase in current beyond a voltage of -0.6 V for both cases corresponds to short-circuit conditions, where some copper dendrites have spanned from cathode to anode, thereby allowing electrons to pass freely. Although the current-voltage response for both PE membranes is similar to that of the CN membranes, minor discrepancies may be observed at a voltage below -0.15 V, where differences in the current output are significant. This is possibly a result of differences in solvent uptake, affected by the extent of membrane wetting by the aqueous solvent, despite the fact that the membranes were soaked in electrolyte overnight before cells were assembled for analysis.

As in other systems with deionization shock waves, it can be more stable to control the current rather than the voltage[28,63], so we perform galvano-electrochemical impedance spectroscopies (GEIS) for PE(+) and PE(-) membranes, in Fig. 7b and 7c, at different direct current biases with alternating currents of 10 µA from 100 kHz to 100 mHz. When applying no dc-bias, the impedance for both cases exhibits a similar response, devoid of any diffusional resistance. When applying a dc-bias, the Warburg-like arc for PE(-) shrinks as the current increases. In contrast, as a result of ion blocking by surface conduction in PE(+), the low frequency response becomes noisy. This may also indicate effects of electro-osmotic surface convection[31,32,50], mostly likely around connected loops in the porous network[40], which could serve to bypass the blocked surface conduction pathways in PE(+) and lead to the observed dendrite penetration. In any case, it is clear that the positive and negative membranes exhibit distinct low frequency responses with increasing dc-bias, which indicates a significant difference in the mass-transfer mechanism for $Cu^{2+}$ associated with the surface charge of the porous medium.

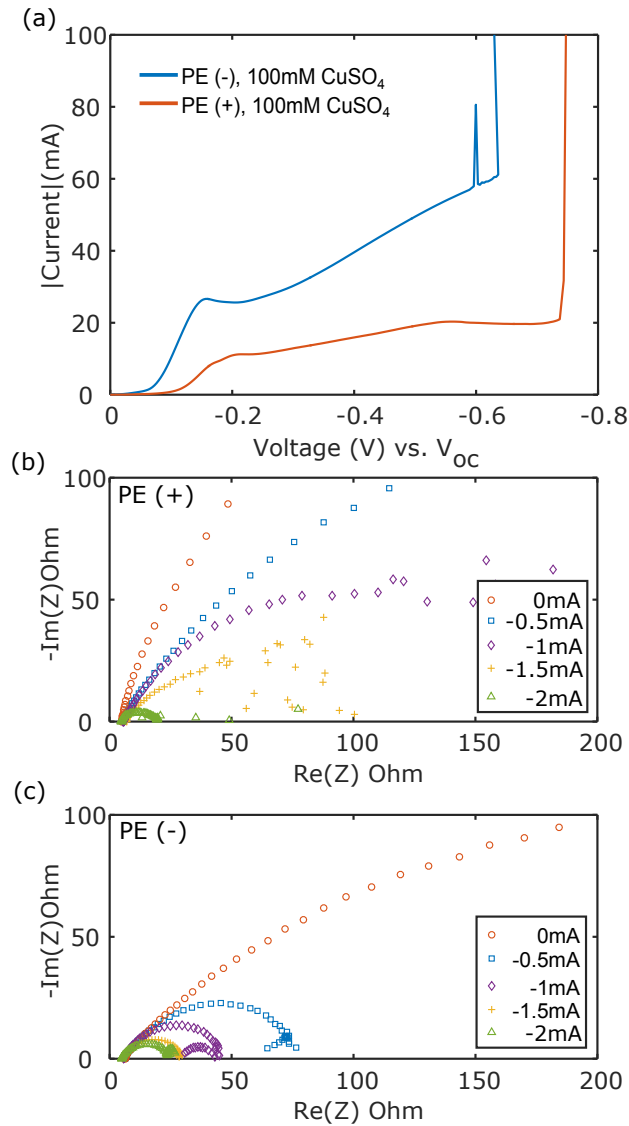

**Figure 7**. Linear sweep voltammetry of Celgard K2045 (a) positive and negative membranes of exposed area 1.327 cm$^2$ between two Cu electrodes in 100 mM $CuSO_4$ at 2 mV/s. Nyquist plot of the Galvano electrochemical impedance spectroscopy of Celgard K2045 (b) positive and (c) negative membranes with the same cell configuration as that of voltammetry.

We observe similar current-voltage response of surface-modified PE membranes in 10 mM $CuSO_4$ solution as those of PE membranes in 100 mM $CuSO_4$ solution (Fig. 8). The nonlinear effect of surface conduction dominants the charge transport as the cathode is polarized beyond -0.15 V. As evident in Fig. 8(a), a sharp difference between the current-voltage behavior of PE(+) and PE(-) membranes further supports the proposition of surface charge sensitivity. Seven copper cells with PE(-) membranes were individual assembled and examined to testify the repeatability of our methodology. As is evident in Fig. 8(b), repeatability can be achieved with stringent LBL-coating procedure as well as cell-assembly process to further validate our proposition of surface conduction phenomenon.

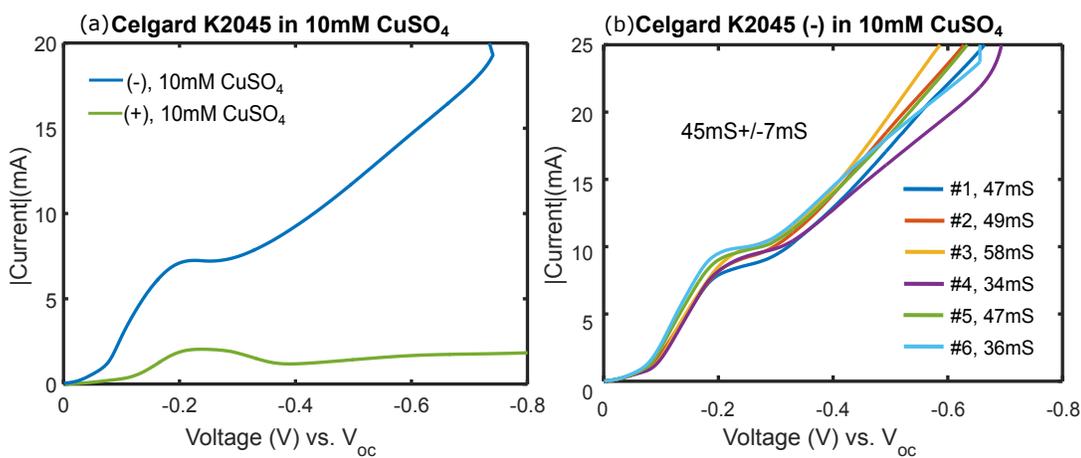

**Figure 8.** (a) LSV of PE(+) and PE(-) membrane between two Cu electrodes in 10 mM $CuSO_4$ solution. (b) LSV of PE(-) membrane between two Cu electrodes in 10 mM $CuSO_4$ solution.

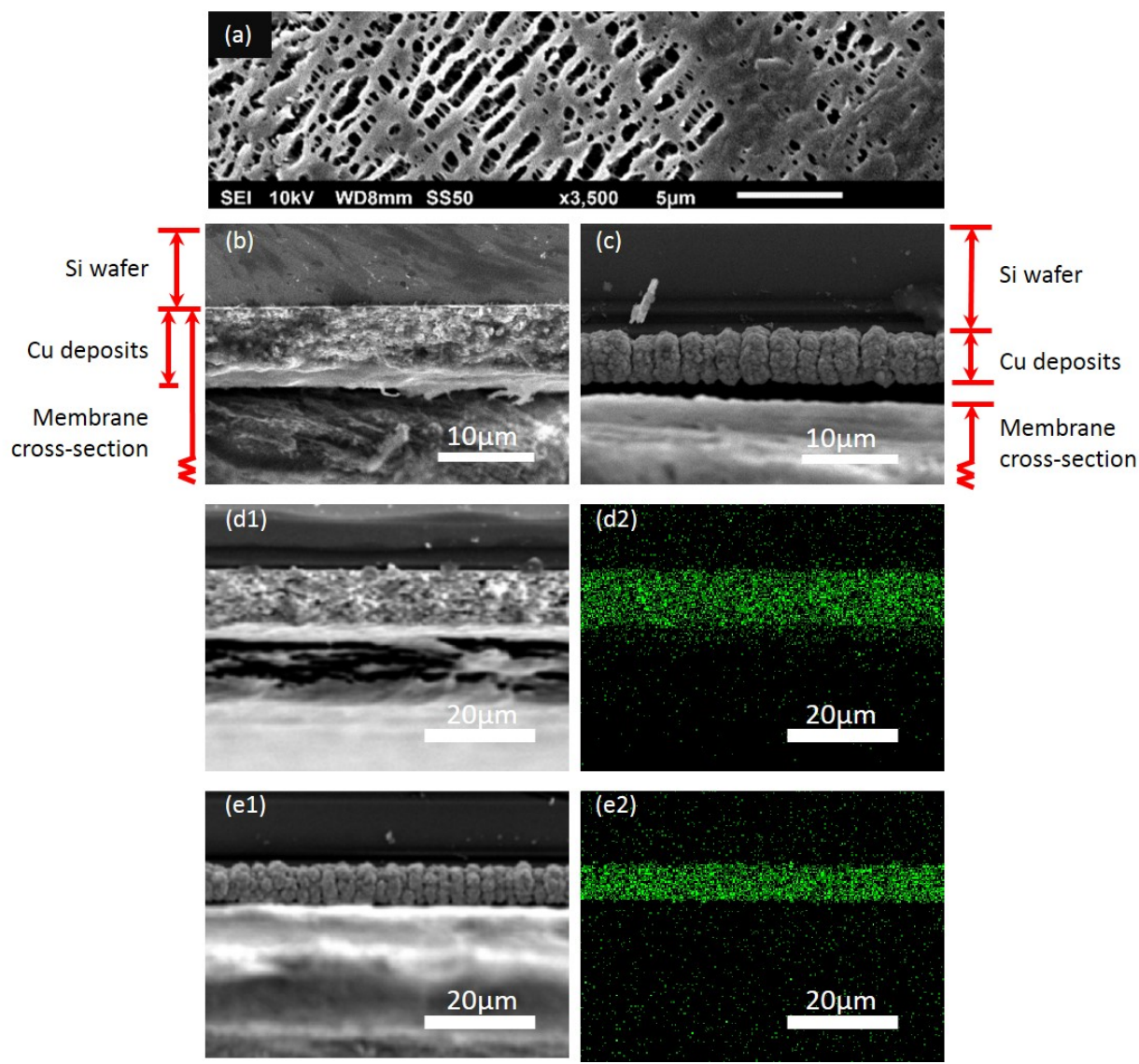

**Figure 9**. (a) SEM image of a random porous polyethylene membrane (Celgard K2045). SEM images of the cross section of a (b) positively and negatively (c) charged polyethylene membrane after chronopotentiometry in 100 mM of $CuSO_4$ at a current of -20 mA for 2000 s. EDS mapping analysis of the corresponding (d) PE(-) and (d) PE(+) membranes. The images before the EDS mapping (d1) and (e1) as well as the corresponding mapping of Cu element (d2) and (e2).

To further support the electrochemical evidence for SC-controlled growth, we performed SEM and EDS mapping analyses to examine the morphological differences between copper electrodeposits in the positive and negative PE membranes. The surface of a random porous membrane before electrodeposition is shown in Fig. 9(a). After 2000 s of galvanostatic deposition of copper onto a silicon wafer (with a thin layer of copper) in 100 mM $CuSO_4$, two distinct cross-sectional morphologies are observed, depending only on the surface charge of the membrane. In the case of PE(-), in Fig. 9(b) and 9(d), a dense copper film (approx. 8 $\mu$m thick) is observed within the membrane. Due to the existence of denser copper within the upper portion of the membrane, the lower portion of the membrane beneath the film, without any copper deposits, is tapered, deformed, and torn away when the cell is disassembled for imaging. In contrast, a layer of porous copper grown directly on the wafer is observed for PE(+), in Fig. 9(c) and 9(e). The whole membrane beneath the copper layer is partially separated from the wafer/copper complex since no copper is deposited into the membrane to provide any adhesion. These morphological discrepancies are consistent with the growth modes described in Fig. 2, where the negatively charged membrane supports the growth of uniform metal-matrix composite, while the positively charged membrane blocks metal penetration.

**CONCLUSIONS**

This work provides fundamental insights into the physics of transport-limited pattern formation in charged porous media. We show that the surface charge and microstructure of porous separators can strongly influence the morphology of copper electrodeposition, which is considered to be the prototypical case of unstable diffusion-limited growth in free solutions. For the first time, we directly observe the suppression of dendritic instability at high rates, exceeding diffusion limitation. With negative surface charge, uniform metal growth is stabilized behind a propagating deionization shock, and reversible cycling is possible. Under the same conditions with positive surface charge, dendrites are blocked from penetrating the medium, and at high rates the growth becomes unstable and cannot be cycled.

Besides its fundamental interest, shock electrodeposition may find applications in energy storage and manufacturing. High-rate rechargeable metal batteries could be enabled by charged porous separators or charged composite metal electrodes[11-13,60]. The rapid growth of dense, uniform metal electrodeposits in charged porous media could also be applied to the fabrication of copper[61] or nickel[10] metal matrix composites for abrasives or wear-resistant coatings.

**METHODS**

**Chemicals:** Polydiallyldimethylammonium chloride (pDADMAC, 100,000 ~ 200,000 $M_w$, 20 wt% in water), (poly(styrenesulfonate) (PSS, 70000 $M_w$), copper sulfate ($CuSO_4$, ≥ 98%), sodium chloride (NaCl, ≥ 98%), and sodium hydroxide (NaOH, ≥ 98%) are purchased from Aldrich and used without further purification. Ultrapure deionized water is obtained from Thermo Scientific (Model No. 50129872 (3 UV)) or from a Milli-Q Advantage A10 water purification system. Cellulose nitrate (CN) membranes (pore diameter 200~300 nm, porosity 0.66 – 0.88, thickness 130 μm, diameter 47 mm) are purchased from Whatman. Polyethylene membranes (K2405) with a pore size of 50 nm, a porosity of 47% and a thickness of 20 μm, are obtained from Celgard. Copper plates (1/8" thickness) were purchased from McMaster Carr and machined down to appropriate dimensions using a water jet cutter.

**Sample Preparation:** The surface charge of CN and PE membranes is modified by layer-by-layer (LBL) method of charged polyelectrolytes. Polydiallyldimethylammonium chloride (pDADMAC) is directly deposited on the membrane to make a positive surface charge, CN(+). For this, the bare CN is immersed in polycation solution (1 mg/mL pDADMAC in 20 mM NaCl at pH 6) for 30 min. Then, the membrane is triple rinsed (10 min each) with purified waterr purification system) to remove unattached polyelectrolyte. Negatively charged CN(-) is obtained by coating negative polyelectrolytes (poly(styrenesulfonate), PSS) on the pDADMAC-coated CN by immersion in a polyanion solution (1 mg/mL PSS in 20 mM NaCl at pH 6) for 30 min and followed by the same washing procedure. The polyelectrolytes coated CN membranes are stored in a $CuSO_4$ solution.

The surface charge of PE membranes are modified using a similar LBL procedures described above. Bare PE membranes are air-plasma treated for 10 min before being immersed in pDADMAC solution for 12 h to make the positively charged membrane (PE(+)). The membrane is triple rinsed (30 min each) with purified water is needed to remove any unattached polyelectrolyte. For the negatively charged PE membrane, thoroughly rinsed PE(+) membranes are immersed in PSS for 12 h, followed by the same washing procedure as that of the PE(+) membrane. The surface-modified PE membranes are stored in purified water and soaked in a $CuSO_4$ solution 12 h before cell-assembly.

**Experiments Apparatus:** The experimental set-up is from previous our work (see Ref. 28). The modified membrane is clamped between two Cu disk electrodes (13 mm diameter) under constant pressure, where Cu is stripped from the anode and deposited on the cathode. Electrode polishing consists of grinding by fine sand paper (1200, Norton) followed by 3.0 µm alumina slurry (No. 50361-05, Type DX, Electron Microscopy Sciences) and thorough rinsing with purified water. For SEM images, a Cu-sputtered Si wafer (1.0 cm × 1.5 cm) is used as a cathode, in place of a copper disk electrode. To prevent the evaporation of the binary electrolyte solution inside the CN or PE membrane, the electrochemical cell is immersed in a beaker containing the same electrolyte. All electrochemical measurements are performed with a potentiostat (Reference 3000, Gamry Instruments). The morphology and composition of electrodeposited Cu films are confirmed by scanning electron microscopy (SEM) with energy-dispersive spectroscopy (EDS) X-ray detector (6010LA, JEOL) at 15 kV accelerating voltage.

**ACKNOWLEDGEMENTS**

This work was supported by Saint Gobain Ceramics and Plastics, Northboro Research and Development Center and by Robert Bosch LLC through the MIT Energy Initiative. J.-H. H. also acknowledges initial support from the Basic Science Research Program of the National Research Foundation of Korea funded by the Ministry of Education (2012R1A6A3A03039224).

**AUTHOR CONTRIBUTIONS**

J.H.H. and M.Z.B. conceived of the idea and designed the experiments. J.H.H. and M.W. prepared the samples. J.H.H., M.W., and P.B. performed electrochemical and SEM analysis. F.R.B. and M.Z.B. supervised the study. J.H.H., F.R.B., and M.Z.B. wrote the manuscript. All the authors discussed the results and commented on the manuscript.

**COMPLETING FINANCIAL INTERESTS**

The authors declare no completing financial interests